\documentclass[prl,amsmath,amssymb]{revtex4}

\usepackage{graphicx}
\usepackage{dcolumn}
\usepackage{bm}

\begin{document}

\title{A correlation between the amount of dark matter in elliptical galaxies
and their shape.}

\author{A. Deur \\
University of Virginia, Charlottesville, VA 22904}

\date{\today}

\begin{abstract}
We discuss the correlation between the dark matter content of 
elliptical galaxies and their ellipticities. We then explore a mechanism 
for which the correlation would emerge naturally. Such mechanism leads 
to identifying the dark matter particles to gravitons. A similar mechanism is 
known in Quantum Chromodynamics (QCD) and is essential to our 
understanding of the mass and structure of baryonic matter.
\end{abstract}

\maketitle

\paragraph{\textbf{Introduction}}

The influence of dark matter is ubiquitous in the universe \cite{Komatsu}.
Let us list but a few cosmological observations which indicate that
the presence of dark matter is necessary:
\begin{itemize}
\item Dark matter explains why the outskirts of disk galaxies can spin so
rapidly \cite{Rubin rot. curves}. At radii greater than about
10 kpc for typical spiral galaxies, rotational speeds are significantly
faster than expected if these galaxies would consist only of baryonic
matter bounded by Newtonian gravity. 
\item Dark matter keeps galaxies confined in clusters even though galactic
speeds exceed the liberation speeds expected from the cluster baryonic
mass and Newton's Law \cite{Zwicky}.
\item Dark matter is necessary to aggregate baryonic matter from its relatively
smooth primordial distribution to the large-scale structures presently
observed \cite{galaxy formation}. 
\item Dark matter solves the abundance problem in the primordial nucleosynthesis
of deuterons \cite{BBN}. 
\end{itemize}
Dark matter is important to understand the internal dynamics of galaxies.
There are many correlations between the different quantities characterizing
a galaxy. The reasons for some of them are understood while others
are still phenomenological observations yet to be explained. The most
evident characteristics of an elliptical galaxy are its mass $M$
and ellipticity $\varepsilon$. From our present understanding of
galaxy formation and galactic dynamics, there is no reason for $M$
and $\varepsilon$ to be significantly correlated. However, dark matter
phenomenology is still not well understood at galactic scale. It is
thus worthwhile to investigate whether the principal properties of
elliptical galaxies are correlated.

\paragraph{\textbf{The correlation between mass and ellipticity}}

Such an investigation was done in Ref. \cite{Deur MNRAS}, with
a significant correlation established. For the study, we have considered
only publications reporting the total mass of at least several elliptical
galaxies. Various selection criteria were applied to isolate the signal
from the background. Only elliptical galaxies of medium masses without
peculiarities were retained. In particular, galaxies in significant
interaction with other galaxies were rejected, as were giant or dwarf
galaxies. The selection criteria were devised to select one type of
elliptical galaxy (typical medium size galaxies) in a relaxed state
so that its mass estimate is reliable. The criteria were determined
before investigating the mass-ellipticity correlation and, as such,
are unlikely to have biased the analysis. In all, 685 determinations
of total galactic masses were used. With such a large number, the
problem of knowing the intrinsic ellipticity of the galaxies (remember
that only projected ellipticities can be observed) can be addressed
statistically. 

Different methods were used in the publications to assess galactic
masses. The analyses were based on the virial theorem \cite{virial},
stellar dynamics modeling \cite{stellar dyn.}, interstellar
gas X-ray emissions \cite{X-ray}, observation of planetary nebulae
and globular clusters \cite{PNe}-\cite{Romanowsky}, observation
of gas disks embedded in elliptical galaxies \cite{gas disks }
or strong lensing \cite{lensing}. The references for each given
method indicate the publications used in the data mining analysis
of Ref. \cite{Deur MNRAS}. The independence of the multiple
techniques guards the analysis \cite{Deur MNRAS} against methodological
biases.

\begin{figure}
\protect\includegraphics[scale=0.55]{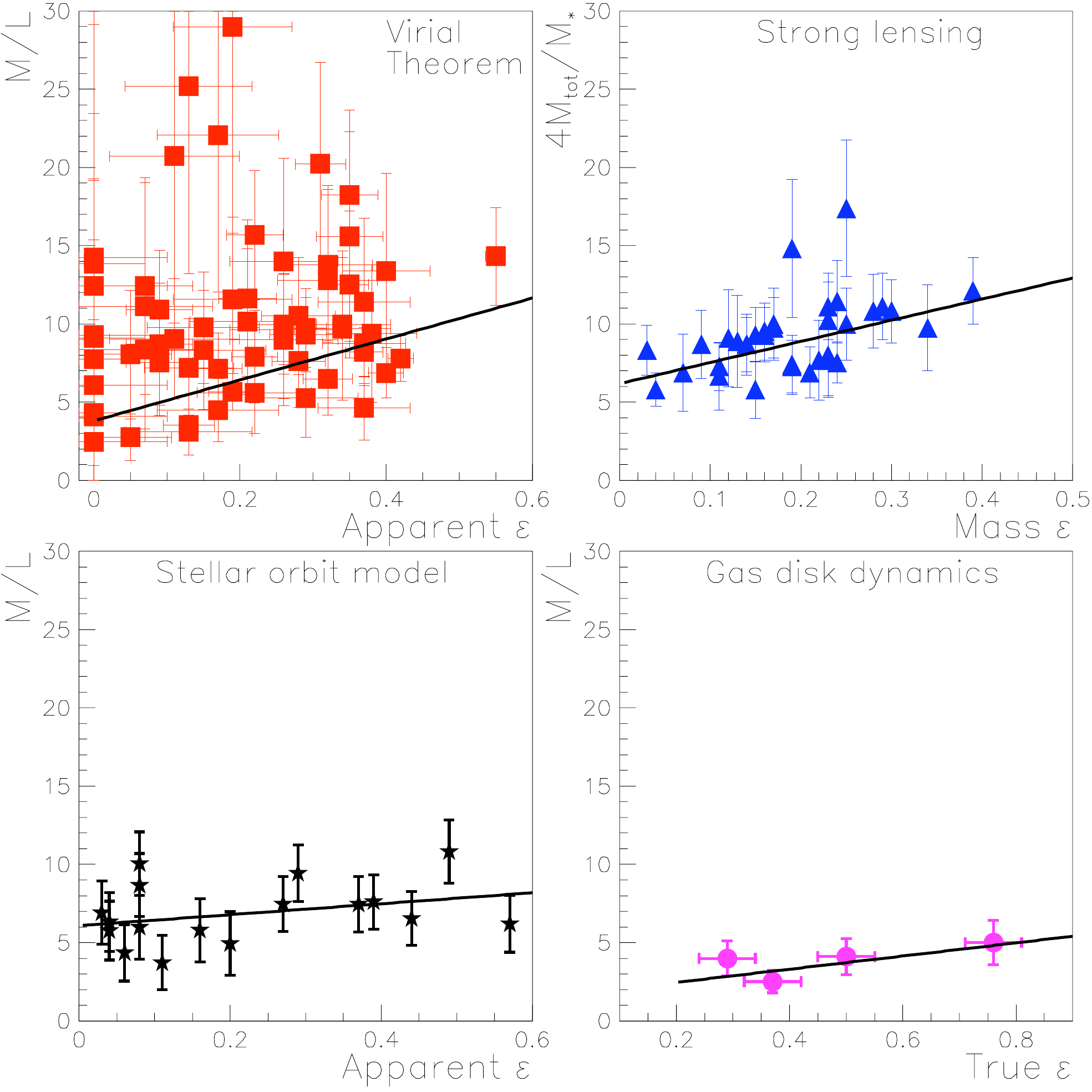}
\caption{\label{fig: sample of results }
Galactic mass over luminosity in solar $\mbox{M}_{\odot}/\mbox{L}_{\odot}$
units versus ellipticity from four publications using different methods
for assessing galactic masses. 
}
\end{figure}

\begin{figure}
\protect\includegraphics[scale=0.55]{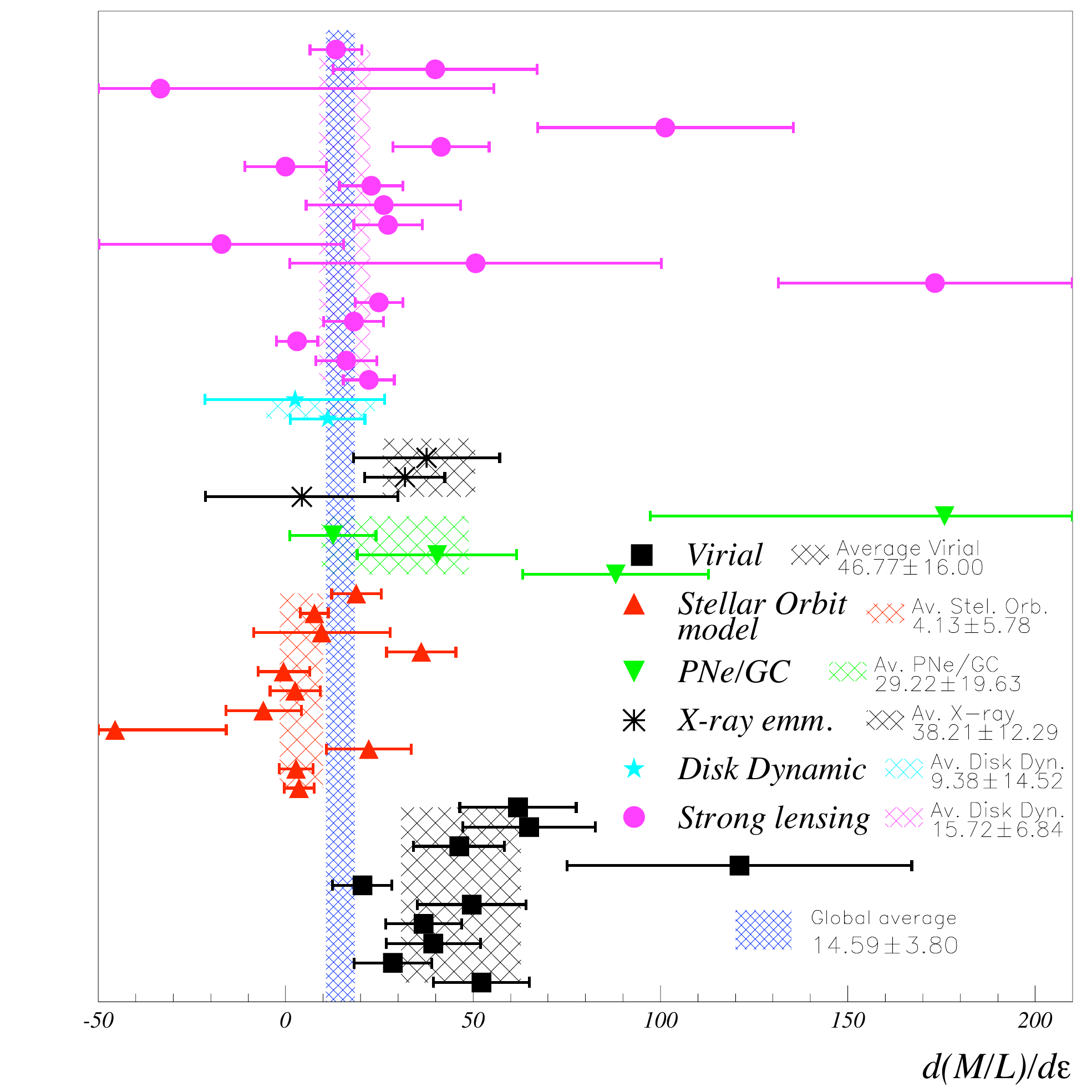}
\caption{\label{fig: Global}
Slopes $\mbox{d}(M/L)/\mbox{d}\varepsilon$ for the samples from Refs.
\cite{virial}-\cite{lensing}. The six different symbols distinguish
the methods used to obtain $M/L$. The corresponding bands indicate
their averages. The blue band is the global average of all methods.
}
\end{figure}

The total masses normalized to galactic luminosity, $M/L$, (or to
stellar mass, $M/M_{*}$) were plotted versus the ellipticities $\varepsilon$.
The relations were then fitted by a straight line for the 41 samples
of galaxies given in \cite{virial}-\cite{lensing}. Four examples
of such fits out of the 41 are shown in Fig. \ref{fig: sample of results }.
 A non-zero slope, $\mbox{d}(M/L)/\mbox{d}\varepsilon\gtrsim3\sigma$
(with $\sigma$ the uncertainty of the slope), signals a correlation
for the given sample. The values of $\mbox{d}(M/L)/\mbox{d}\varepsilon$
were then corrected on a statistical basis for the fact that we only
observe the projection of the ellipse with an unknown viewing angle.
To assess the correction, we assumed that the galaxies are axisymmetric,
and thus can be characterized by a single ellipticity, $\varepsilon_{true}$.
A Gaussian distribution of $\varepsilon_{true}$ was generated. Assuming
that galaxies are randomly oriented with respect to Earth, the simulated
distribution of projected  --- or apparent ---  ellipticities $\varepsilon_{app}$
is obtained from the initial distribution of $\varepsilon_{true}$.
The parameters of the Gaussian distribution were then adjusted until
the simulated distribution of $\varepsilon_{app}$ matched the observed
one. Setting $M/L=a\varepsilon_{true}$, $a$ was determined by matching
the constructed two-dimensional distribution of $a\varepsilon_{true}$
vs $\varepsilon_{app}$ to the experimental distribution of $M/L$
vs. $\varepsilon_{app}$. The value of $a$ gives the slope $\mbox{d}(M/L)/\mbox{d}\varepsilon$,
corrected for the projection effect. The 41 corrected determinations
of $\mbox{d}(M/L)/\mbox{d}\varepsilon$ were combined, accounting
for their respective statistical accuracies, their precision, different
$M/L$ normalizations, and correlations between mass determinations
using the same technique and overlapping sets of galaxies. Fig. \ref{fig: Global}
shows the 41 corrected determinations of $\mbox{d}(M/L)/\mbox{d}\varepsilon$,
the average for each method and the overall average. %
 All averages indicate a positive slope $\mbox{d}(M/L)/\mbox{d}\varepsilon$,
although some are compatible with zero. The overall average $\mbox{d}(M/L)/\mbox{d}\varepsilon=(14.59\pm3.80)$M$_{\odot}$/L$_{\odot}$
is clearly positive. This slope is steep given the average $M/L=7.7$
M$_{\odot}$/L$_{\odot}$. 

The correlation between $M/L$ and $\varepsilon$ can be physical
or it could be a systematic bias. A methodological bias is unlikely
since data from independent methods and different authors were used.
Other possible biases were investigated. The correlation was re-assessed
by using the Pearson correlation coefficient and by fitting the data
with a quadratic rather than a linear form. These re-assessments also
indicate a strong correlation. A survival analysis was performed in
which individual data sets were removed in turn and the correlation
re-estimated. The resulting distribution of the $\mbox{d}(M/L)/\mbox{d}\varepsilon$
has a root-mean square width of 0.514, well below the average value
of 14.59. The possibility of lenticular (S0) galaxy contamination
of the samples was investigated, since S0 and elliptical galaxies
are difficult to distinguish at small $\varepsilon$. Since the $M/L$
ratios of S0 galaxies tend to be smaller than those of elliptical
galaxies, a contamination would induce an increase of $M/L$ with
$\varepsilon$. However, even the upper limit for S0 contamination
was found to be too small to explain the observed correlation. The
correlation could also originate from the many relations between quantities
describing galaxies. This was investigated and no interrelations leading
to a ($M/L$)-$\varepsilon$ correlation could be identified. Thus,
at the present state of our knowledge, the correlation seems to be
physical rather than originating from a methodological, observational
or instrumental bias. In any case, the correlation, regardless of
its origin, is an important fact that must be considered in studies
of galaxy dynamics.

The average correlation slope $\mbox{d}(M/L)/\mbox{d}\varepsilon=(14.59\pm3.80)$M$_{\odot}$/L$_{\odot}$
and the average value $M/L=7.7$ M$_{\odot}$/L$_{\odot}$ correspond
to the intercept $M/L=(3.3\pm1.5)$M$_{\odot}$/L$_{\odot}$ at $\varepsilon=0$.
With the normalization used here, corresponding to luminosities obtained
in the blue band, $M_{*}/L\simeq4$ M$_{\odot}$/L$_{\odot}$ when
only the stellar mass $M_{*}$ is counted. Thus, the roundest galaxies,
including those considered in \cite{Romanowsky}, contain little
dark matter. The correlation is puzzling since there is no basic reasons
for it in the context of the Cold Dark Matter model and since dark
matter is needed to trigger the formations of structures that will
latter form elliptical galaxies.

\paragraph{\textbf{Possible Explanation: Gravitons as dark matter particles}}

Considering gravitons as the particles that constitute dark matter
offers a possible solution to these puzzles. (A correlation between
$M/L$ and $\varepsilon$ was predicted in this context \cite{Deur PLB}.)
At galactic distance scales and for the gravity fields involved, the
classical formulation of gravity, General Relativity (GR), can be
used to investigate the behavior of gravitons and to determine if
they can account for dark matter. Consequently, although we will be
using the language of quantum field theory and discuss gravitons rather
than classical fields, the argument does not depend on a contingent
quantization of gravity. 

Since gravitons carry energy and momentum, they couple to each other
with a coupling $\sqrt{G}$ at the amplitude level, where $G$ is
the Newton's constant. This well-known fact is described e.g. in textbooks
\cite{Feynman/Kiefer}-\cite{Salam/Zee} and, in classical field
language, is responsible for the non-linearity of GR's equations.
In natural units ($\hbar=c=1$), the coupling $G$ is small, e.g.
the gravitational potential near a proton is $\frac{GM_{p}}{r}=3.8\times10^{-38}$,
with $M_{p}$ the proton mass and $r=8.4\times10^{-16}$ m the proton
radius \cite{proton radius}, to be compared with the force couplings
$\alpha\approx7\times10^{-3}$ for the electromagnetic force (QED)
and $\alpha_{s}\approx1$ for the strong nuclear force (QCD). The
extreme smallness of $G$ makes the effects of mutual coupling of
gravitons usually negligible. However, for massive enough systems,
the effect should become important. For example, for a typical galaxy,
$\frac{GM}{r}\simeq10^{-3}$. 

The Lagrangian of GR is $\mathcal{L}_{GR}=\frac{1}{16\pi G}\sqrt{det(g_{\mu\nu})}g_{\mu\nu}R^{\mu\nu}$,
with $g_{\mu\nu}$ the metric tensor and $R_{\mu\nu}$ the Ricci tensor.
Expanding $\mathcal{L}_{GR}$ in terms of the tensor gravity field
$\varphi_{\mu\nu}$ yields \cite{Salam/Zee}:

\begin{eqnarray}
\mathcal{L}_{GR}={\left[\partial\varphi\partial\varphi\right]+}\sqrt{16\pi G}\left[\varphi\partial\varphi\partial\varphi\right]+16\pi G\left[\varphi^{2}\partial\varphi\partial\varphi\right]+\ldots+\sqrt{16\pi G}\,\varphi_{\mu\nu}T^{\mu\nu},\label{eq:Einstein-Hilber Lagrangian}\end{eqnarray}
where $T^{\mu\nu}$ is the energy-momentum tensor and $\left[\varphi^{n}\partial\varphi\partial\varphi\right]$
is a shorthand notation for a sum over the possible Lorentz invariant
terms of the form $\varphi^{n}\partial\varphi\partial\varphi$. For
example, $\left[\partial\varphi\partial\varphi\right]$ is explicitly
given by the Fierz-Pauli Lagrangian \cite{Fierz-Pauli}, the
first order linear approximation of GR that leads to Newton's gravity
in the case of static ($v\ll c$) bodies:

\begin{eqnarray}
\left[\partial\varphi\partial\varphi\right]=\frac{1}{2}\partial^{\lambda}\varphi_{\mu\nu}\partial_{\lambda}\varphi^{\mu\nu}-\frac{1}{2}\partial^{\lambda}\varphi_{\mu}^{\mu}\partial_{\lambda}\varphi_{\nu}^{\nu}-\partial^{\lambda}\varphi_{\lambda\nu}\partial_{\mu}\varphi^{\mu\nu}-\partial^{\nu}\varphi_{\lambda}^{\lambda}\partial^{\mu}\varphi_{\mu\nu}.\label{eq:Fierz-Pauli Lagrangian}\end{eqnarray}
Although the coupling $\sqrt{16\pi G}$ is small, the terms $\left(16\pi G\right)^{n/2}\left[\varphi^{n}\partial\varphi\partial\varphi\right]$
may become important for large enough $\varphi_{\mu\nu}$, i.e. massive
enough bodies. We will first explore the phenomenology of Eq. (\ref{eq:Einstein-Hilber Lagrangian})
and then discuss how to quantitatively study its consequences, following
a numerical method developed In Ref. \cite{Deur PLB}. 

The polynomial expansion Eq. (\ref{eq:Einstein-Hilber Lagrangian})
allows to interpret $\mathcal{L}_{GR}$ in the language of particle
physics. The first term $\mathcal{L}{}_{GR}^{1}\equiv\left[\partial\varphi\partial\varphi\right]$
generates the free graviton propagator, producing in the static limit
Newton's law with its familiar $1/r^{2}$ dependence. Higher order
terms $\mathcal{L}{}_{GR}^{n}\equiv\left(16\pi G\right)^{n/2}\left[\varphi^{n}\partial\varphi\partial\varphi\right]$
represent graviton vertices with $n+2$ external legs: the gravitons
interact with each other. For a massive enough static two-body system,
the gravitons are preferably attracted toward the region of highest
graviton density, i.e. the line joining the two bodies. This and the
fact that the two bodies are static render the two dimensions transverse
to this line irrelevant. The system is reduced to one dimension. In
one dimension, a force mediated by massless carriers is constant.
Thus, the attraction between the two bodies becomes constant rather
than varying as $1/r^{2}$ as in the free-propagation case. The binding
is stronger, effectively leading to an increase of the system mass.
This increase contributes to a dark mass. For a homogeneous continuous
disk distribution, the gravitons are attracted to the disk plane.
In the extreme case, the propagation of gravitons is confined in two
dimensions, resulting in a $1/r$ dependence of the gravity of the disk
acting on a mass within the disk. For a spherically symmetric system
there is no preferred direction(s) and the force from the sphere acting
on one of its constituents varies as $1/r^{2}$. Thus, the mutual interaction
of gravitons and the symmetry of the matter distribution could explain
the correlation between $M/L$ and $\varepsilon$. Namely, the interaction,
or the total effective mass of the system, varies from the familiar
$1/r^{2}$ law for $\varepsilon=0$ systems to a $1/r$ law for $\varepsilon=1$
systems. The question is whether elliptical galaxies are massive enough
to make the terms $\mathcal{L}{}_{GR}^{n}$, $n>1$, non-negligible.
We now discuss quantitatively this question.

\begin{figure}
\protect\includegraphics[scale=0.55]{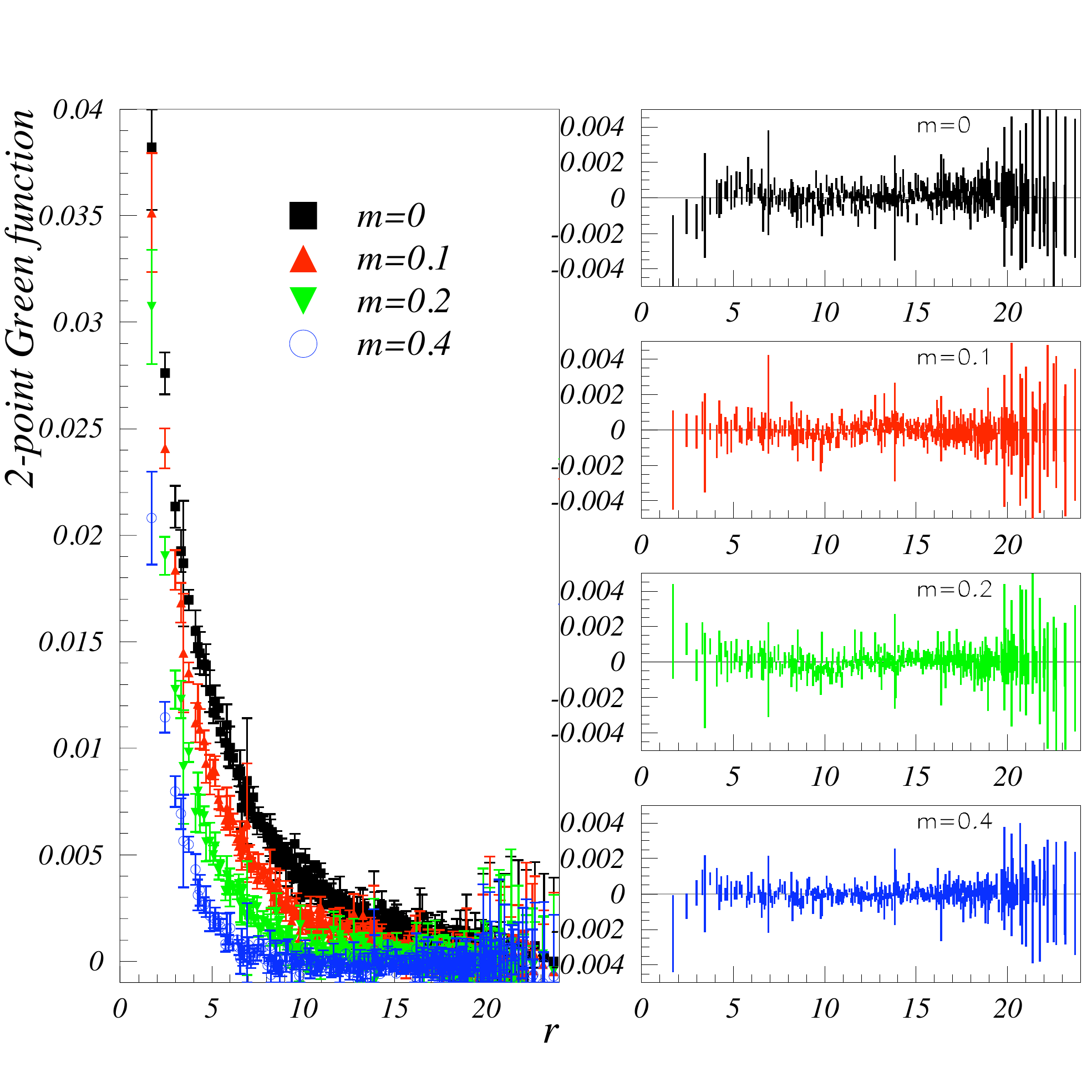}
\caption{\label{fig: abelian case}
Potentials obtained by Monte-Carlo calculations of the gravitational
field when the higher order terms $\mathcal{L}{}_{GR}^{n}\equiv$$\left(16\pi G\right)^{n/2}\left[\varphi^{n}\partial\varphi\partial\varphi\right]$
in $\mathcal{L}_{GR}$ are ignored and a fictitious mass term $m^{2}\varphi^{2}$
is added (left). Newton's case corresponds to $m=0$. Yukawa potentials
are obtained when $m\neq0$. Residuals from the $e^{-mr}/r$ expectation
are show on the right.
}
\end{figure}

\begin{figure}
\protect\includegraphics[scale=0.50]{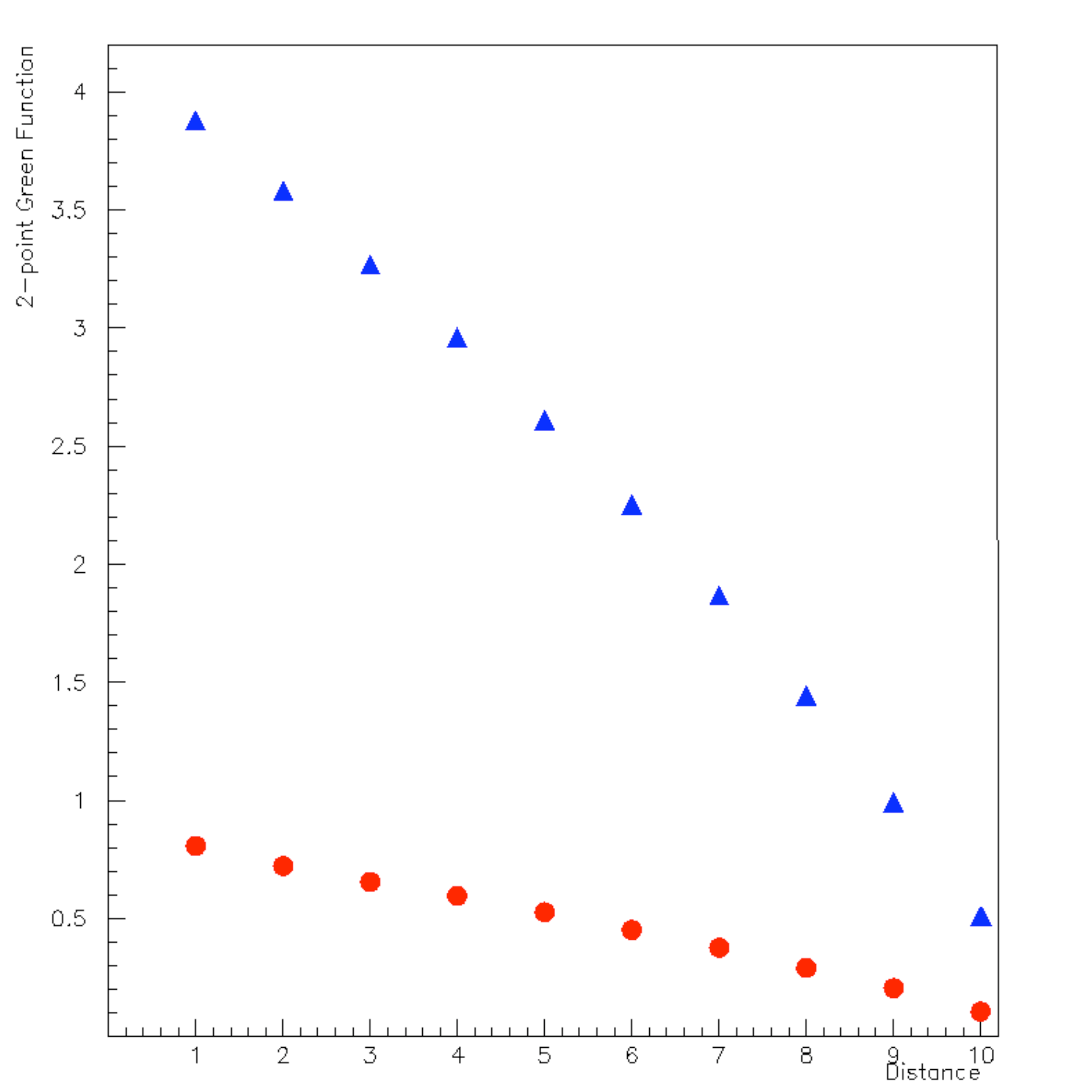}
\caption{\label{fig: non-abelian case}
Potentials obtained when the terms $\mathcal{L}{}_{GR}^{n}$, $n=1,$
2, in $\mathcal{L}_{GR}$ are included. The potentials are now linear.
The red circles (blue triangles) are for a 2-body system, each body
of mass $10^{12}$M$_{\odot}$ ($3.2\times10^{12}$M$_{\odot}$) typical
of elliptical galaxies. 
}
\end{figure}

For the static case of two point-like bodies located at $r_{1}$ and
$r{}_{2}$, the tensors in Eq. (\ref{eq:Einstein-Hilber Lagrangian})
can be approximated by their time-time components \cite{Deur PLB}
and $\mathcal{L}_{GR}$ becomes:

\begin{eqnarray}
\mathcal{L}_{GR}=\sum_{n=0}^{\infty}a_{n}\left(16\pi G\right)^{n/2}\varphi^{n}\partial\varphi\partial\varphi+\sqrt{16\pi G}\,\varphi\left(\delta^{(4)}(r-r_{1})+\delta^{(4)}(r-r_{2})\right),\label{eq:Scalar Lgrangian}\end{eqnarray}
where $\varphi\equiv\varphi_{00}$ and $a_{n}$ are coefficients comparable
to unity. For order of magnitude estimates, we can take $a_{n}=1$
for all $n$. (One can show that $a_{1}=1$ by deriving the potential
by using Eq. (\ref{eq:Scalar Lgrangian}) for weak fields and comparing
it to the Einstein\textendash{}Infeld\textendash{}Hoffmann equations
\cite{Einstein Infled Hoffmann}.) In the static case, the potential
is given by the two-point Green function $G_{2p}(r)$. In the Feynman
path-integral formalism it is: 

\begin{eqnarray}
G_{2p}(r_{1}-r_{2})=\frac{1}{Z}\intop D\varphi\,\varphi(r_{1})\varphi(r_{2})e^{-iS},\label{eq:2pt green}\end{eqnarray}
where $S\equiv{\int d^{4}x\mathcal{L}_{GR}}$ is the action, $\intop D\varphi$
sums over all possible field configurations, and $Z=\intop D\varphi\, e^{-iS}$.
In the static case, the time dimension can be ignored. The gravity
field $\varphi$ can then be numerically calculated at each site of
a 3D lattice by using the standard Metropolis Monte-Carlo method.
The method can be tested for known cases. Ignoring the terms $\mathcal{L}{}_{GR}^{n}$,
$n>1$, leads to the expected Newtonian potential, $G_{2p}(r)\propto1/r$.
Another check can be done by adding a fictitious graviton mass term
in the Lagrangian, $m^{2}\varphi^{2}$. This leads to the expected
Yukawa potential, $G_{2p}(r)\propto e^{-mr}/r$. These results are
shown in Fig. \ref{fig: abelian case}. %
Including the terms $\mathcal{L}{}_{GR}^{n}$ with $1\leq n\leq2$
yields roughly linear potentials in the case of a system of two bodies
of typical galaxy mass, $M\sim10^{12}$ M$_{\odot}$, see Fig. \ref{fig: non-abelian case}.
These calculations indicate that galaxies are in the regime in which
the terms $\mathcal{L}{}_{GR}^{n}$, $n>1$, become important. Symmetry
arguments imply that the effect decreases when the system becomes
more and more spherically symmetric. Consequently, the total effective
mass of galaxies should correlate with its ellipticity, with little
dark matter in the roundest systems. The symmetry arguments also imply
that for disks with baryonic mass densities decreasing exponentially
with radius, as for disk galaxies, rotation curves should reach a
plateau \cite{Deur PLB}, a well known manifestation of dark
matter \cite{Rubin rot. curves}. The calculations in Ref. \cite{Deur PLB}
also agree well with galaxy cluster dynamics and the Bullet Cluster
observations \cite{bullet and DM}.

\paragraph{\textbf{What can QCD teach us about dark matter?}}

Gravity is not the only force that involves self-interacting fields.
The theory of the strong interaction of quarks and gluons, QCD, is
the archetypical self-interacting field theory. It is consequently
worthwhile to explore the parallels between gravity and QCD. Both
are Yang-Mills (non-Abelian) theories for which the symmetry group
is non-commutative. The underlying reason is that the gauge charges
in both theories (the color charges for QCD and the energy-momentum
tensor for gravity) are matrices, and thus non-commuting quantities.
The physical consequence is that the force carriers (gluons for QCD
and gravitons for gravity) mutually interact. Phenomenologically,
the interactions occur because gluons carry color charges and gravitons
have energy-momentum. Mathematically, the field Lagrangians for QCD
and gravity have a similar form, although with different Lorentz structures:
gluons have spin 1, so QCD is a vector field; gravitons have spin
2, so gravity is a tensor field. The QCD Lagrangian, without the matter
term, is:

\begin{eqnarray}
\mathcal{L}_{QCD}=\psi_{\mu\nu}^{a}\psi_{a}^{\mu\nu},\label{eq:Pure field Lagrangian, QCD 1}\end{eqnarray}
with $\psi_{\mu\nu}^{a}=\partial_{\mu}\psi_{\nu}^{a}-\partial_{\nu}\psi_{\mu}^{a}-\sqrt{4\pi\alpha_{s}}f_{abc}\psi_{\mu}^{b}\psi_{\nu}^{c}$,
where $\psi_{\mu}^{a}$ are the gluon fields, $a$ the gluon color
indices and $f_{abc}$ the SU(3) structure constants. Expanding $\mathcal{L}_{QCD}$
leads to

\begin{eqnarray}
\mathcal{L}_{QCD}= & [\partial\psi\partial\psi]+\sqrt{4\pi\alpha_{s}}[\psi^{2}\partial\psi]+4\pi\alpha_{s}[\psi^{4}],\label{eq:QCD pure field lagrangian 2}\end{eqnarray}
with the explicit structure $[\partial\psi\partial\psi]=2\partial_{\mu}\psi_{\nu}^{a}\partial^{\mu}\psi_{a}^{\nu}-2\partial_{\mu}\psi_{\nu}^{a}\partial^{\nu}\psi_{a}^{\mu}$,
$[\psi^{2}\partial\psi]=2f_{abc}\left[\left(\partial_{\mu}\psi_{\nu}^{a}\right)\psi_{b}^{\nu}\psi_{c}^{\mu}-\left(\partial_{\mu}\psi_{\nu}^{a}\right)\psi_{b}^{\mu}\psi_{c}^{\nu}\right]$
and $[\psi^{4}]=f_{abc}f_{ade}\psi_{\mu}^{b}\psi_{\nu}^{c}\psi_{d}^{\mu}\psi_{e}^{\nu}$.
This is to be compared to the first three terms of $\mathcal{L}_{GR}$
in Eq. (\ref{eq:Einstein-Hilber Lagrangian}). The comparison reveals
the similarity in the structure of $\mathcal{L}_{QCD}$ and $\mathcal{L}_{GR}$.
There are, however, significant differences. One is that $\sqrt{G}$
is small while $\alpha_{s}$ is large for distances typical of hadron
sizes ($\sim$10$^{-16}$ m). Another difference is that the gravity
field is spin-2 and hence always attracts, while the strong force
field is spin-1 and can either attract or repulse. That $\sqrt{G}$
is small can be compensated by the fact that gravity always attracts.
It is then reasonable to expect, for massive enough systems, that
gravity's self-interaction should yield effects similar to those characterizing
the hadron structure. Such effects in hadron structure phenomenology
are: 
\begin{itemize}
\item The strength of QCD becomes large at long distances (quarks are confined).
The accepted explanation (in the static case of heavy quarks) is that
$\alpha_{s}$ is large and gluons are color-charged. The gluon flux
between two quarks collapses into a flux tube inducing a string-like
confining potential. 
\item For a family of hadrons, the square of the hadron mass $m_{H}$ is
proportional to the angular momentum $J$ ({}``Regge trajectories''):
$\log\, m_{H}=0.5\,\log\, J+c$, where $c$ is a constant. The interpretation
is that larger $J$ imply larger centripetal forces and hence, to
keep the quark system bound, larger string tension (i.e. binding energy):
the more a hadron rotates, the larger its total mass. 
\item There are no strong interactions outside hadrons, except for small
residual effects such as the force resulting from light hadron exchange
(Yukawa forces). This is because gluons, as color-charged particles,
are confined into hadrons as well.
\end{itemize}
We can confront this list to the following cosmological phenomena:
\begin{itemize}
\item The strength of gravity in galaxies or more massive systems is larger
than expected, based on the observed amount of visible matter. The
accepted explanation of the increase in gravity's strength is the
existence of additional, non-baryonic, matter.
\item The luminosities of disk galaxies or, equivalently, their luminous
masses $M_{*}$, are related to their maximum rotation speed $v$
(Tully-Fisher relation): $\log\, M_{*}=3.9\,\log\, v+1.5$. The faster
a galaxy rotates, the larger its total mass.
\item Dark energy effectively acts as negative pressure, currently balancing
the effect of matter's attraction on the expansion of the universe. 
\end{itemize}
There is an intriguing correspondence between the two lists. It is
tempting to attribute it to the similarity between the Lagrangians
of gravity and QCD. Such origin of dark matter and dark energy would
yield natural explanations of the dark matter-baryon coincidence \cite{Baryon-DM coincidence}
and the cosmic coincidence problems \cite{cosmic coincidence}.
It would also explain the negative results for experimental searches
of WIMP \cite{LUX} and axion dark matter candidates \cite{Axion }.

\paragraph{\textbf{Summary}}

We discussed a strong correlation between the dark matter content
of elliptical galaxies and their ellipticities. It implies that the
roundest elliptical galaxies contain little dark matter, a puzzling
fact in the context of galaxy formation and cold dark matter models.
The mutual interaction of gravitons suggests an explanation. It also
directly explains the flat rotation curves of disk galaxies and cluster
dynamics. Such observations can be paralleled with QCD phenomenology.
The similar forms of the field Lagrangians of gravity and QCD may
explain the observed correspondences.

\paragraph{\textbf{Acknowledgments }}

The author thanks Haiyan Gao for inviting him at Duke University and
G. Cates, F.-X. Girod, J. Gomez, C. Mu\~{n}oz-Camacho, A. Sandorfi,
S. \v{S}irca and B. Terzi\'{c} for useful discussions on the work
reported here.

\end{document}